# W.W. MORGAN AND THE DISCOVERY OF THE SPIRAL ARM STRUCTURE OF OUR GALAXY

**William Sheehan**
*2105 Sixth Avenue SE, Willmar, MN 56201, USA.*
E-mail: sheehan41@charter.net

*In memoriam*—Donald Edward Osterbrock (1924-2007)

**Abstract:** William Wilson Morgan was one of the great astronomers of the twentieth century. He considered himself a morphologist, and was preoccupied throughout his career with matters of classification. Though his early life was difficult, and his pursuit of astronomy as a career was opposed by his father, he took a position at Yerkes Observatory in 1926 and remained there for the rest of his working life. Thematically, his work was also a unified whole. Beginning with spectroscopic studies under Otto Struve at Yerkes Observatory, by the late 1930s he concentrated particularly on the young O and B stars. His work on stellar classification led to the Morgan-Keenan-Kellman [MKK] system of classification of stars, and later—as he grappled with the question of the intrinsic color and brightness of stars at great distances—to the Johnson-Morgan UBV system for measuring stellar colors. Eventually these concerns with classification and method led to his greatest single achievement—the recognition of the nearby spiral arms of our Galaxy by tracing the OB associations and HII regions that outline them. After years of intensive work on the problem of galactic structure, the discovery came in a blinding flash of Archimedean insight as he walked under the night sky between his office and his house in the autumn of 1951. His optical discovery of the spiral arms preceded the radio-mapping of the spiral arms by more than a year. Morgan suffered a nervous breakdown soon after he announced his discovery, however, and so was prevented from publishing a complete account of his work. As a result of that, and the announcement soon afterward of the first radio maps of the spiral arms, the uniqueness of his achievement was not fully appreciated at the time.

**Keywords:** W.W. Morgan, spiral arms of our Galaxy, radio astronomy, Jan Oort, Yerkes Observatory, Edwin Frost, Otto Struve, stellar classification, MKK system, OB stars, OB associations, Walter Baade, Andromeda Nebula, HII regions, Jason Nassau, Stewart Sharpless, Donald Osterbrock, Local Arm, Perseus Arm, pattern-recognition, the nature of scientific discovery.

"The things for this year are a deeper view and understanding of the character and depth of the mind and the completion of the first stage of the work on the evolution of galaxies. There is room for a more ordered picture in both areas; how beautiful are the vistas in each region—how beautiful and deep they are! And how similar are the aesthetic consideration and world-laws in both! The work of art; the world of the galaxies; the world of the mind; all – ALL – in the fundamental world of forms." (W.W. Morgan, New Year's resolutions, 1957).

## 1 INTRODUCTION

It is now common knowledge that our Galaxy is a vast spiral star system which we view edgewise from within one of the spiral arms. The first clear demonstration of the fact, however, by Yerkes Observatory astronomer William Wilson Morgan (Figure 1), occurred only in 1951. This was one of the grandest discoveries in the history of astronomy, and when Morgan presented it, in a fifteen minute talk at the American Astronomical Society meeting in Cleveland the day after Christmas 1951, he received a resounding ovation, that included not only clapping but stomping of feet.[1] But for various reasons—not least that Morgan suffered a nervous breakdown that led to hospitalization within months after the discovery—no definitive account of his discovery appeared at the time (but see Anonymous, 1952, and Morgan, Sharpless and Osterbrock, 1952: 3).

Morgan had used optical methods to detect the nearer spiral arms. When he left Billings Hospital at the University of Chicago, Morgan was determined to reconstitute himself and reorganize his psyche through a systematic program of self-help and psychoanalysis which he would document in a remarkable series of personal notebooks he kept for most of the rest of his life. By the time he returned to the Yerkes Observatory, Jan Oort and his Dutch and Australian collaborators had independently announced the discovery of the spiral-arm structure of our Galaxy on the basis of radio astronomical observations. At the time their results seemed more far-reaching, since whereas Morgan had only identified the nearby spiral arms, the radio astronomers were able to identify structures on the hidden far side of the Galaxy.

For a time the discoveries of Oort and his collaborators overshadowed Morgan's work. Only later, in about 1970, was it realized that their distances were not as accurate as had been supposed because of large-scale systematic deviations from circular motion of the hydrogen gas clouds on which they had relied for their maps; thus, the radio maps turned out not to be very reliable, and the uniqueness of Morgan's achievement began once more to be fully appreciated (see Burton, 1976: 279-281).

## 2 THE KAPTEYN UNIVERSE AND THE STRUCTURE OF OUR GALAXY

At the beginning of the twentieth century, the standard view of the Galaxy was that offered by the Dutch astronomer J.C. Kapteyn (Kruit and Brekel, 2000) who, much as William Herschel had done in 1785 in proposing his 'grindstone model of the universe', still regarded the Galaxy as a small disk of stars. Since Kapteyn ignored the effects of extinction of starlight by interstellar dust, his model, like Herschel's, included only the nearer stars; as a result, his disk, measuring a mere 4,000 parsecs long by 1,000 parsecs wide, remained centered on or near the Solar System. It was this smallish star-system that George Ellery Hale set out to investigate—along lines defined by Kapteyn—with the 60-inch reflector at Mt. Wilson Observatory (which saw 'first light' in 1908). Kapteyn spent the summers between 1909 and 1914 as a Research Associate at Mt. Wilson, and was Hale's most influential adviser.[2]

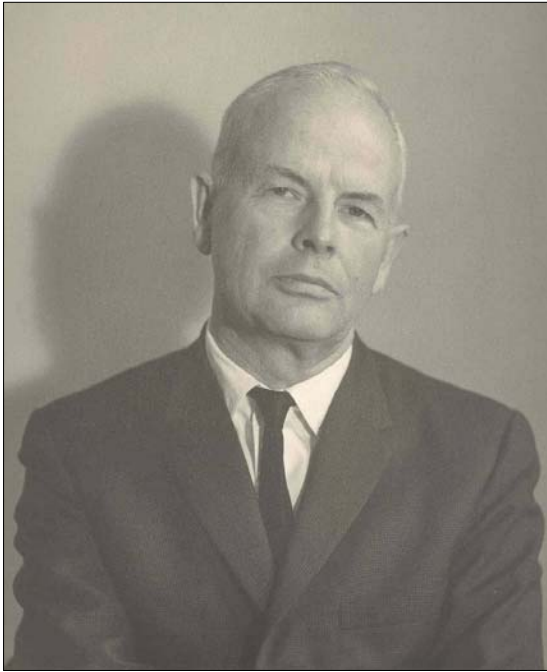

Figure 1: W.W. Morgan, at about the time he first documented the existence of spiral arms in our Galaxy (courtesy: Yerkes Observatory Archives).

Kapteyn's views about the structure and size of the Galaxy were eventually undermined by the work of Harlow Shapley, who also used the Mt. Wilson 60-inch reflector. In 1914 Shapley (then a 29-year-old Princeton Ph.D.) began making photometric measurements of the stars in globular and galactic clusters, little suspecting that this line of investigation would ultimately lead to unlocking the Sun's position in the Galaxy—to what he rather colorfully would call the 'galactocentric revolution'. He correctly surmised that extinction of starlight by interstellar dust was negligible in directions away from the Galactic Plane; however, his subsequent observations of stars in galactic clusters led him to erroneously extend that result to the Galactic Plane itself. And yet though he made mistakes, he was able, in 1917, to establish the main result. "In [his sixth paper]," wrote Allan Sandage (2004: 300), "… Shapley invented three powerful and (it turned out) highly reliable methods to determine cluster distances. It was a singular achievement." Armed with these tools and his 60-inch plates (as well as earlier plates taken by Solon Bailey at Harvard's Southern Station at Arequipa), Shapley succeeded in mapping the distribution of the globular clusters and showed that they form a framework located eccentrically to the Sun. Thus, Shapley deduced that the Sun was far removed from the center of the Galaxy and that the latter was much larger than Herschel, Kapteyn or anyone else had imagined. In the end, Shapley's model displaced the smaller, Sun-centered Kapteyn Universe.

Meanwhile, it was becoming clear, especially from the wide-angle Milky Way photographs taken by the American astronomer E.E. Barnard (Sheehan, 1995), that the dark regions in the Milky Way were not tubules or holes perforating a disk of stars, as earlier astronomers such as William and John Herschel had supposed, but dust clouds scattered along the Galactic Plane.

By this time many astronomers believed that the so-called 'spiral nebulae' were extragalactic star systems. Many had been discovered by the Herschels, while Lord Rosse and his assistants with the great reflector at Birr Castle, Ireland, had discerned a spiral structure in a number of them, most famously in the 'Whirlpool Nebula' in Canes Venatici, but also in eighty or so others. Faint and small spirals were later found by the thousands in deep plates taken by James Keeler with the 36-inch Crossley Reflector at Lick Observatory at the end of the nineteenth century, and appeared to be virtually numberless in the regions around the Galactic Poles (Osterbrock, 1984).

Although Keeler himself leaned toward the view that these spirals were planetary systems in formation, a later Lick astronomer, Heber D. Curtis, who studied the Crossley images more carefully, discerned a family resemblance in all the spiral nebulae; in other words, they appeared to form a class of similar objects that were distributed at different angles and at different distances. In each case where they were seen edge-on, dark rifts divided them, which Curtis recognized as similar to the dark dust clouds of the Milky Way that Barnard had photographed. By 1917, the same year Shapley published his paper on the globular clusters, Curtis was arguing that the spirals were star systems, or 'island universes', a result that seemed to receive confirmation with the discovery of novae in the spiral nebula NGC 6946 by G.W. Ritchey, again using the 60-inch reflector at Mount Wilson, followed by Curtis's discovery of additional novae in other spirals (Osterbrock, 2001a). Shapley himself remained unconvinced, and was misled by his overestimate of the size of the Galaxy into supposing that the spiral nebulae must be local, and also by his (and everyone else's) failure to grasp the sheer violence of the supernova explosion which had occurred in the Andromeda Nebula in 1885,[3] thus setting the stage for the famous Curtis-Shapley debate of 1920. The conclusive demonstration that the Andromeda Nebula was an extragalactic star-system finally came with Edwin Hubble's discovery of its Cepheid variables using the 100-inch Reflector on Mt. Wilson in 1923-1924; even Shapley accepted the implications at once.

Within a little more than a decade, Kapteyn's rather quaint model of the Galaxy had been completely discredited, largely owing to the pioneering work of the 60-inch telescope at Mt. Wilson, and our Galaxy became one of countless millions of 'star systems' strewn throughout the Universe. It might well be a majestic spiral in its own right (as suggested as early as 1900 by the Dutch amateur, Cornelis Easton), though it might also be a flattened elliptical. Determining its actual form proved to be one of the most daunting problems of twentieth-century astronomy.

## 3 TWENTIETH CENTURY STUDIES OF GALACTIC STRUCTURE

Kapteyn died in June 1922, but largely as a result of his influence Dutch astronomers continued to work on galactic structure. Among them were his student P.J. van Rhijn at the Kapteyn Institute in Groningen, Jan Oort (Figure 2) at Leiden, and Bart J. Bok (who studied under Oort at Leiden, received his doctorate from Groningen, and then did much of his work at Harvard).

The first to propose a rotating model of the Galaxy was the Swedish astronomer Bertil Lindblad. However, finding Lindblad's mathematical treatment impenetrable, Oort decided to devise his own approach to the problem. Realizing that there was much more interstellar extinction by dust than had been realized, Oort surmised that the best way of understanding galactic structure would be to study the motions of the stars and introduced the concept that because of galactic rotation there was a well-defined relationship between the radial velocities, distances and angles of stars in a rotating system. This meant that "… measured systematic radial velocities could be converted to approximate distances in a straightforward way." (Osterbrock, 2001b: 147). This was a very important idea that would underpin his later investigations into the structure of our Galaxy.

Unfortunately the Netherlands is one of the worst places imaginable for observational astronomy, and Oort did not have the telescopes to provide the kinds of data he needed. But he learned of the discovery of radio emission from the Galaxy by the American engineer, Grote Reber (Reber's first paper was published in 1940), and grasped the great potential of a new and powerful technique which would allow penetration of the interstellar dust clouds. He thus posed an important question to H.C. van de Hulst, a brilliant student in Utrecht who had written a noteworthy paper on interstellar dust: "Is there a spectral line at radio frequencies we should in principle be able to detect? If so, because at radio frequencies extincttion would be negligible, we should be able to derive the structure of the Galaxy. We might even be able to detect spiral arms, if they exist." (Katgert-Merkelijn, 1997).

After several months of study Van de Hulst found that there was indeed such a spectral line, the 21-cm line of the ground state of hydrogen (neutral hydrogen, HI). Given the vast abundance of hydrogen in the gas in the Galactic Plane, Oort at once realized that mapping of the interstellar atomic hydrogen would likely led to the discovery of our Galaxy's spiral arms. Other advantages of the radio technique lay in the fact that it gave very high velocity (frequency) resolution—which meant that the results could be immediately tested against Oort's rotation model for the Galaxy.

Unfortunately, this work was delayed by World War II, and afterwards there were further delays in getting the proper equipment. A particularly frustrating setback occurred when one of Oort's receivers was destroyed in a fire. Nevertheless, Oort persisted, and in collaboration with the radio engineer C.A. Muller he finally succeeded in detecting the 21-cm line with an antenna at Kootwijk, on 11 May 1951, just six weeks after the feat had been accomplished at Harvard by Edward M. Purcell and H.I. 'Doc' Ewen (for details see van Woerden and Strom, 2006). (In contrast to the financially-strapped Dutch, the Americans had been beneficiaries of a crash wartime radar research program.) Oort and his Dutch and Australian colleagues now began working on a systematic study of the structure of our Galaxy, including hitherto inaccessible regions on the far side of the Galactic Center. This seemed to make the mapping of the spiral arms by radio astronomers inevitable; indeed, the first such maps appeared within a year. But the radio astronomers were forestalled by the narrowest of margins, as it was optical astronomers who were the first to achieve this result.

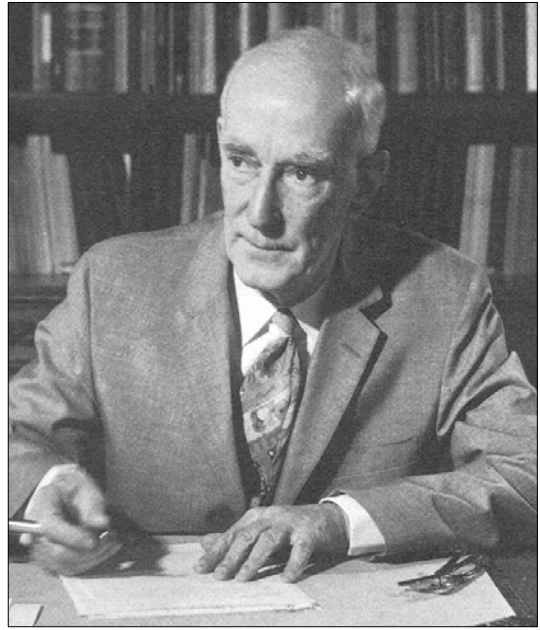

Figure 2: Jan Oort, 1900-1992 (after Oort, 1981: frontispiece).

For a long time, optical astronomers had been tackling the problem of galactic structure by means of brute-force star counts and methods of statistical analysis, such as those developed by Kapteyn. The basic idea was simple: as one counted stars, the number of stars would rise in the vicinity of a spiral arm and then drop off beyond it. No one applied these methods more diligently than Oort's former student Bok, but after ten years of hard slogging he had failed to find any expected stellar density concentrations that could be identified with spiral arms. By the late 1930s, he thought the task of tracing out the spiral structure of our Galaxy was almost hopeless: he later recalled that "In public lectures during that period I often said it was unlikely the problem would be solved in my lifetime." (cited in Croswell, 1995: 74).

Clearly a different approach was needed. By the middle of the twentieth century, optical astronomers had largely realized this and had regrouped around the idea of using the most luminous stars in their mapping efforts. The final breakthrough came when William Wilson Morgan, building on the brilliant work of Walter Baade at Mt. Wilson, put together a number of leads during the late 1940s and early 1950s and forged, from what others might have perceived as unrelated scraps, a technique that dramatically revealed the hitherto unglimpsed spiral arms of our Galaxy.

## 4  W.W. MORGAN: THE MAN[4]

Like two other important figures in galactic research, E.E. Barnard and Carl Seyfert, Morgan was a native of Tennessee. He was born on 3 January 1906 in the tiny hamlet of Bethesda, which no longer exists. His parents were William Thomas Morgan and Mary (née Wilson) Morgan, Southern Methodist Church home missionaries. During the first part of his professional career, Morgan always wrote his name as 'W.W. Morgan' rather than 'William Morgan', presumably in order to establish an identity that was independent from that of his father.

During W.W. Morgan's childhood, the Morgan family was constantly on the move in the South of the USA, and until the age of nine he was entirely home-schooled with his mother. A list of all the places he lived, recorded on a scrap of paper in the Yerkes Observatory Archives (Morgan, n.d.), shows that from Bethesda he moved to Crystal River, Florida, in December 1908; to Starke, Florida, in 1910, where he saw Comet 1P/Halley; to Punta Gorda in 1912; to Key West in December of that same year; to a farm 18 miles from Punta Gorda the following year; to Perry, Florida, in the latter part of 1914; to Colorado Springs, Colorado, in December 1915; to Poplar Bluff, Missouri, in October of the following year; to Spartanburg, South Carolina, in the summer of 1918; to Washington, D.C., the following summer; to Fredericktown, Missouri, in September 1919; and back to Washington in the spring of 1921.

The foregoing paragraph furnishes some insight into the basis of what became Morgan's obsession: a quest for permanence, the need to achieve a firm foothold and, above all, to find what the poet John Keats (1818: 302) once called "… certain points and resting places." In geographical terms, it would lead to his well-known clinging to Yerkes Observatory, where he lived and worked for more than sixty-eight years until his death in 1994. But perhaps just as significant was his need for conceptual fixed and secure resting places, which led to his attempt to develop a system of stellar classification that would be secure, and would not be overturned as a result of later revisions to the calibrations, as previous schemes of classification had been.

All his life, Morgan was haunted by his relationship with his domineering and unstable father, who seems to have been a man of great energy but who was also moody and given to dogmatic, intolerant views. He was possibly manic-depressive; there was no doubt he was sometimes emotionally and physically abusive toward his family, including Morgan. In later years, Morgan may have exaggerated the extent of the abuse; however, he did rather vividly recall being almost beaten to death at the age of two, only to be saved by the timely intervention of his mother. In an interview recorded in May 1993 (by which time he was suffering quite advanced Alzheimer's Disease) Morgan claimed that he was "… beaten up frequently …" by his father (see Croswell, 1995: 75).

As with others who have suffered from unhappy and abusive childhoods, Morgan found a refuge in the stars. In an interview he recalled:

> The stars gave me something that I felt I could stay alive with. The stars and the constellations were with me, in the sense that on walks in the evening, I was a part of a landscape which was the stars themselves. It helped me to survive. (Croswell, ibid.).

His father, William T. Morgan, seems to have started out as a fire-and-brimstone preacher who took a rigid and literal-minded interpretation of the Scriptures like that associated with the Scopes monkey-trial in Dayton, Tennessee. He interpreted the prohibition against working on the Sabbath literally, so that when young Morgan was in school he was forbidden to do any work on Sundays at all. As a result he was always falling behind:

> I remember late in high school, in Washington, D.C., I always dreaded Sunday night because I never was prepared for Monday. So it was a question of just survival. Just passing was all. And that's what it was like through these years. (Morgan, 1978).

Whereas William T. had become the same kind of man as his own father, a coal-miner from Warrior, Alabama, William W., the future astronomer, went about forming his personality-structure by what Freud called *reaction-formation*: the process of psychologically defining the self in opposition to, and outside of, the problematic person's perspective rather than by identifying with it. His father, who was awe-inspiring, powerful, capricious and terrifying, eventually found a career as an itinerant inspirational speaker and was absent for long periods of time. He wanted Morgan to follow in his footsteps, but Morgan wisely recognized that he did not have the same kind of personality as his father and that he would never be happy in such a role.

Morgan's first formal encounter with astronomy was "… as a refuge from an unhappy childhood, during the Influenza epidemic in the winter of 1918-19." (Morgan, 1987). His father had left for an extended period of time, and Morgan, with his mother and sister, moved to Frederickstown, Missouri, where a Methodist junior college (Marvin College) and an attached high school were located in a cow pasture. Morgan entered high school there in the fall of 1919. He received his first astronomy book (a collection of star maps) from his Latin teacher, Alice Witherspoon, and she also arranged his first view through a telescope; it was of the Moon. Morgan (ibid.) later recalled the benevolent Miss Witherspoon's decisive influence on his development:

> In addition to the astronomical introductions, she presented the Latin language as a living thing, and helped me to realize that even a "dead" language could possess a vibrant, living form. My preoccupation with morphology (the science of form) probably began with this experience.

At the same time, Morgan discovered his father's set of the Harvard Classics—'Dr. Eliot's six-foot shelf of books', so-called because Harvard President, Charles W. Eliot, had selected them. One of these included the Elizabethan play *Doctor Faustus*, by Christopher Marlowe, and sixty-seven years later Morgan (ibid.) recalled the electrifying effect this had on him:

> The picture of the partially legendary Faustus, the man who longed to press outward toward the horizons of knowledge – and beyond to the stars – has been the ruling passion of [my] life.

Morgan finished his last two years of high school at Central High School in Washington D.C., then in the fall of 1923 he enrolled at Washington and Lee University in Lexington, Virginia. Although he was interested in astronomy, he had no idea at the time that this would become his profession. Instead, he decided to specialize in English, in preparation for a teaching career. However, he performed well in mathematics, physics and chemistry, and even talked his physics teacher, Benjamin Wooten, into acquiring a small astronomical telescope, so that he could observe sunspots.

In the summer of 1926, a year before Morgan was to finish his degree, Wooten made a summer trip to Yerkes Observatory. When he went up the stairs and rang the doorbell, the Director, Edwin Brant Frost, happened to be just inside. Wooten told Frost about the student who had pressed him to buy a telescope. It turned out that Frost was looking for an assistant to operate the Observatory's spectroheliograph and obtain daily images of the Sun, as the previous incumbent, Philip Fox, had just left to take a Chair at Northwestern University. Morgan was offered the job, but there were still difficulties; not least was the fact that Morgan's father was violently opposed, thinking that he would "… end up just in a laboratory working for somebody else, [and] that's nothing." (Morgan, 1978). That was the last time that Morgan talked to his father about anything; his father, who had been absent for long periods previously, now decided to abandon the family completely—Morgan never saw him again, and afterwards could not even find out what had happened to him or the year in which he died.

The rage and disappointment that Morgan felt towards his father would be reflected above all in one symbolic act. After his father left, Morgan appropriated his Harvard Classics, a coveted possession to anyone who valued literature and great ideas. Morgan savagely ripped his father's name plates out of all of the books bar one (where he missed the name plate because it had been accidentally affixed to the back rather than to the inside front cover of a book, like all the rest). One senses that with this act Morgan was symbolically attempting to tear his father out of his life.[5]

Though his father was gone, his image continued to cast a long shadow over Morgan's development. As a result of that problematical relationship, he always feared leaning on others and being 'devoured' by them. He was a sensitive, lonely introvert, who struggled with low self-esteem and feared being hurt by others. He wrote, characteristically, in a note from 1943: "Everything – objects and people – are shadows enduring for an instant. No one can come inside where I am. Where I live nothing can touch me." (Morgan, 1943: 25 July). He added further comments on his problematical relationships with his colleagues in one of his personal notebooks:

> January 5, 1957. Oort has stood for a number of years as a partial father figure for me … Until a few years ago I was afraid of men in general; they seemed to be a superior race to me and to women – with whom I felt much more at ease than with men … In the years 1953-7 I have made real friends – almost for the first time in my life. Of course, I did have friends earlier; but there was a sort of unstable equilibrium in connection with them because of my dependency leanings …
>
> January 19, 1957. 1 A.M. in bed. I am shedding [Bengt] Strömgren for the same general reasons that Freud shed Fliess: because I have been dependent on him; and I plan not to be dependent in that way anymore. (Morgan, 1957a).

Under the circumstances, Morgan was extremely sensitive to criticism, had difficulty feeling accepted by his professional peers, and often felt like an outsider even in his own family. After his breakdown, he entered into the only completely open and entirely comfortable relationship he ever really had, with his 'Dear Book', as he called the personal notebooks he kept compulsively for thirty years; these became his indispensable and constant companion, a confidante which was physically present at all times. It is clear that he personified these little volumes. They were the 'person' who was interested in everything in which he was interested, a sounding board which could be trusted to listen but never talk back and provided him with understanding and acceptance without qualification.[6]

## 5 THE MOVE TO THE YERKES OBSERVATORY

Yerkes Observatory Director, Edwin Brant Frost, had been born with congenital myopia, a condition predisposing to retinal detachments, and he was legally blind by the time Morgan arrived at the Observatory. (He once told Morgan, probably quite seriously, that the immediate cause of his blindness was the strain of correcting the young Edwin Hubble's first scientific paper!).[7] He was a humane and well-rounded sort of person who held great lawn parties (see Figure 3), but scientifically was not very productive. Nevertheless, in that pre-pension era, he tenaciously hung on until retirement age, courageously making his way everyday to his office from Brantwood, his residence, by means of a guy-wire strung along the footpath through the woods.

At first, Morgan lived in the basement of the Observatory (not in the often-unheated and sometimes damp attic-area known as the 'Battleship' because of its porthole-like windows), but soon after his arrival he married Helen Barrett, the daughter of Yerkes astronomer Storrs Barrett, and the couple moved into their own house, 100 yards east of the Observatory (Figure 4); Morgan would remain there for the rest of his life.

In 1931, Frost retired, and he was replaced by Otto Struve, a Russian immigrant from a very distinguished family of astronomers (e.g. see Batten, 1988). Struve (Figure 5) was an imposing, hulking man whose eyes were not quite congruent and who had a gruff, bearish manner. He was incredibly hard-working, and in those early days was a tremendous inspiration and father-figure to Morgan. Struve once remarked that he had never looked at the spectrum of a star, any star, where he did not find something important to work on. The remark made a lasting impression on Morgan and helped set the direction of his career. Although Struve was an astrophysicist and was interested in using stellar spectra as a tool to understand what was going on physically in stars, Morgan was temperamentally drawn to problems of stellar classification. As early as 1935 he produced his first paper on the subject, "A descriptive study of the spectra of A Stars." (Morgan, 1935).

## 6 MORGAN, THE YOUNG SUPERGIANT SLAYER

According to Struve (1953: 282), it was a series of lectures given by Bok at Yerkes one year later that first inspired Morgan "… to improve the distances of the hotter stars and to investigate the structure of the Milky Way with the help of these distances." These hotter stars included stars of spectral type B and their even brighter, but much rarer cousins, the O stars. None of the closer stars are B stars, and there are only a few of them within a distance of 300 light years. But because these stars are so intrinsically bright, they "… dominate the naked-eye sky all out of proportion to their true population." (Kaler, 2002: 183). The B stars include such admirable specimens as Rigel, Achernar, Beta Centauri, Spica, Alpha and Beta Crucis, and Regulus.

The B stars are young hot stars, prominent in the ultraviolet (Figure 6). They are rapidly rotating stars, and in some of them the velocities of rotation can be as high as 200 km/sec, so they eject matter into equatorial rings that radiate emission lines, characteristic of the so-called Be emission stars (Figure 7). They were first grouped together in a spectral classification in the Henry Draper Catalog, which was developed from spectra examined at Harvard College Observatory by Wilhelmina Fleming, Antonia Maury and Annie Jump Cannon, under the supervision of Edward C. Pickering. The Henry Draper Catalog introduced the familiar categories OBAFGKM, and was a one-dimensional classification which, as was eventually proved at Mt. Wilson in 1908, was keyed to temperature. In the hotter stars—those of classes A, B, and O—the spectral type is determined largely by the strength of the hydrogen lines (Balmer lines) and the increasingly dominant presence of lines of singly or doubly ionized helium (in O stars, these include not only absorption but also emission lines).

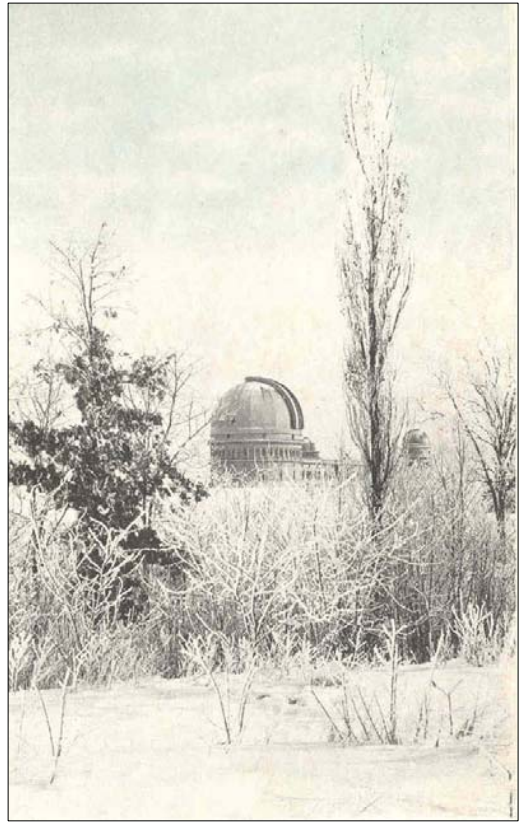

Figure 4: A picturesque postcard view of the Yerkes Observatory in the 1920s (courtesy: Yerkes Observatory Archives).

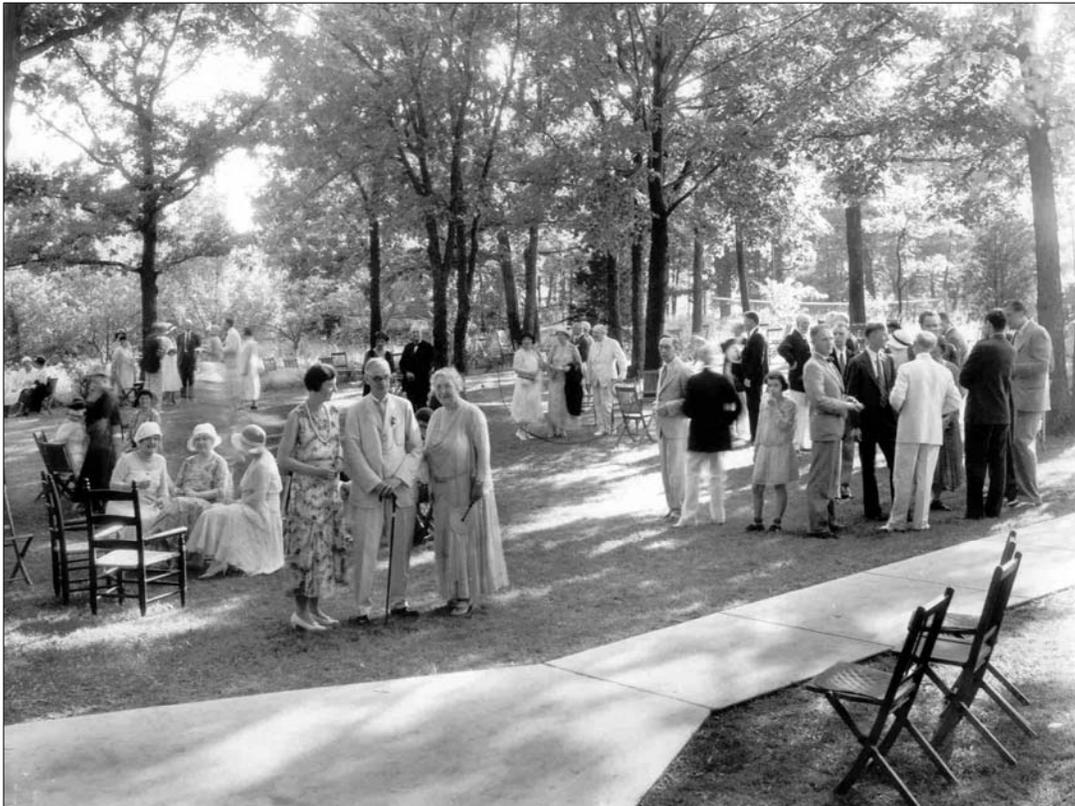

Figure 3: Yerkes Observatory garden party; Edwin Frost, with the cane, is in the foreground just left of centre, flanked by two women (courtesy: Yerkes Observatory Archives).

As early as 1897 Miss Maury had realized that there were distinctly different spectra for stars of a given temperature. In some stars of a given type—Miss Maury's c stars—the hydrogen lines were sharper, while in others they appeared broadened and more diffuse (showing 'wings'). Between 1905 and 1907, the stars in which the lines were sharp were shown by the Danish astronomer, Ejnar Hertsprung, to be much more luminous than the corresponding Main Sequence stars; in other words, they were supergiants. The stars whose lines are broadened into wings are those of the Main Sequence—dwarfs; the wings are produced by the effects of surface gravity and pressure.[8]

The Harvard catalog published by Annie Jump Cannon was based on the way the spectra appeared to her. When good high-resolution spectra of stars began to be obtained with the Mt. Wilson 60-in Telescope,[9] there was much more fine structure visible than had been the case in the Harvard spectrograms. Beginning in 1914, W.S. Adams and A. Kohlschütter at Mt. Wilson began to document in great detail the effects of luminosity on line strengths and line ratios in the spectra of these stars. It turns out that these effects are very sensitive to the precise physical conditions in stars and their atmospheres (as Cecelia Payne-Gaposchkin used to say, all stars, at high enough resolution, appear 'peculiar'). Naturally, the Mt. Wilson astronomers wanted to work out their own classification system in order to deal with all this additional level of detail they were finding, and they went on to develop the first two-dimensional classification system combining temperature and luminosity criteria. Of course, the classes they assigned differed markedly from those Annie Jump Cannon had assigned; inevitably, this produced some tension between the Harvard and Mt. Wilson groups.

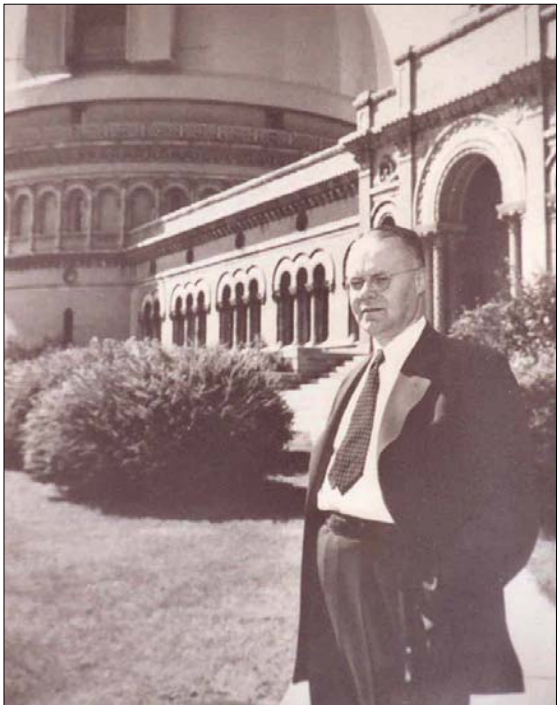
Figure 5: Otto Struve posing outside the Yerkes Observatory (courtesy: Yerkes Observatory Archives).

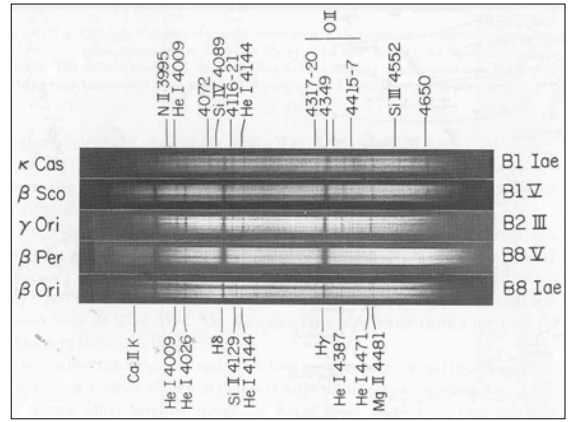
Figure 6: Spectra of five B stars in the MKK Atlas.

The O and B stars were especially problematic. They have weak spectral lines, and because of the great distances of these stars, the Harvard astronomers often had difficulty seeing them at all. Thus, Annie Jump Cannon had classified some heavily-reddened O and B stars as A or even as F. In 1936, when Morgan was beginning to work on spectral classification, he shifted his interest to these high-luminosity stars because they were precisely those where, as Donald Osterbrock notes,

> … the Harvard classification was so bad. Before Morgan, people were using spectral types out of the Henry Draper Catalogue that were not very good. If you take the spectral types as published in the HD and try to use them today, they're terrible. (cited in Croswell, 1995: 78).

The leaders in the classification of high-dispersion spectra of stars at this time were the groups at Mt. Wilson and the Dominion Astrophysical Observatory in Victoria (British Columbia), both of whom tried to relate spectral type to absolute values of luminosity. But when Morgan turned to their presumably more reliable classifications he found discrepancies, because the two groups had adopted different calibrations for their luminosities. It began to appear that the whole field of spectral classification might remain forever in a state of flux and confusion.

Morgan wanted to find a way around this. In the end, he decided to take "… the drastic step of abandoning the assignment of numerical values to absolute luminosities." (Sandage, 2004: 254). With his colleagues Philip Keenan and Edith Kellman (the latter providing clerical assistance), he began working on a new classification scheme. It was a huge undertaking, and was finally published in 1943 as the *Yerkes Atlas of Stellar Spectra, with an Outline of Stellar Classification*. It has became known as the MKK (and more recently as the MK) Atlas.

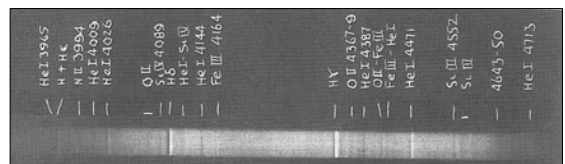
Figure 7: Spectrum of a Be star, taken from the MKK Atlas.

Instead of the continuous absolute magnitude numbers of the Mt. Wilson continuum, Morgan and Kee-nan sorted the Mount Wilson Main Sequence into discrete bins forming five luminosity classes: Ia and Ib supergiants, II bright giants, III giants, IV subgiants and V dwarfs (later adding VI, subdwarfs). The Mt. Wilson absolute magnitude numbers were based on measures of the intensities of specific spectral features. Inevitably, as new measures were made, the luminosity and hence classification of stars underwent constant recalibration and revision.

Morgan wanted his classification system to be secure—true for all time. In the MKK system, peculiar and exceptional stars were set aside. Instead the atlas emphasized 'ordinary stars', Main Sequence stars whose spectra were obtained using the same dispersion, depth of exposure on the photographic plate and method of development (which affects the contrast of bright and dark features). These stars were important statistically as they were the only ones that were suitable for large-scale studies of galactic structure. Morgan's strategy was to choose from among these ordinary stars a series of what he called 'specimens', standard stars defining what he later called a 'box' or reference frame; all other normal stars could then be classified by comparing them to these standard stars. C.R. O'Dell has described Morgan's method:

> The astrophysics was kept in the background. Morgan didn't directly try to relate the morphological spectral features to stellar temperature, luminosity, or gravitational effects at all. As far as the classification scheme was concerned, there was a sequence of boxes each having stars of a particular set of spectral signatures. The adjacent boxes held stars of similar but distinguishably different spectra. The astrophysicist could then come along and interpret these in terms of physical characteristics such as temperature, luminosity, and gravity. (Pers. comm., 31 March 2007).

Since only temperature (or color equivalent) and luminosity were needed to uniquely locate a star's spectrum, Morgan claimed that *by simple visual inspection* of a spectrogram these parameters could be determined and the star placed relative to the comparison stars. There was no need to measure anything. There would be no need—with new sets of measurements of the features of a spectrum, such as line width or intensity—to reshuffle the spectral classifications, since no quantitative value was put on any spectral feature. Morgan vigorously defended this qualitative approach. He admitted it was qualitative; but this did not mean it was indefinite or indeterminate. As he argued:

> The indefiniteness is … only apparent. The observer makes his classification from a variety of considerations – the relative intensity of certain pairs of lines, the extension of the wings of the hydrogen lines [Balmer lines], the intensity of a band – even a characteristic irregularity of a number of blended features in a certain spectral region. To make a quantitative measure of these diverse criteria is a difficult and unnecessary undertaking. In essence the process of classification is in recognizing similarities in the spectrogram being classified to certain standard spectra [those of standard stars]. (Morgan, Keenan and Kellmann, 1943: 4).

In a sense, spectral classification now became a true art-form and required, as Harvard historian of science, Peter Galison (1998: 340), points out,

> … the subjective, the trained eye, and an empirical art, an 'intellectual approach', the identification of 'patterns', the apperception of links 'at a glance', the extraction of a 'typical' sub-sequence with wider variations.

These were skills that defied simple or mechanistic algorithms; the judgments were far too complex for that. Though none of the features in the spectra that Morgan identified as the basis of his classifications is easily quantified—the line ratios (the relative intensities of lines in the spectra of different stars) are extremely variable when they differ appreciably from unity, and the appearance of spectra change greatly with resolution (Andrew T. Young, pers. comm., 14 January 2007)—Morgan insisted that the human eye-brain system (or at any rate *his* eye-brain system) is remarkably adept at just such pattern-recognition tasks. It excels in the discernment of similarities not unlike those involved in recognition of faces.[11] In his introduction to the MKK Atlas, Morgan concluded by specifically calling attention to the analogy between spectral classification and facial recognition tasks:

> It is not necessary to make cephalic measures to identify a human face with certainty or to establish the race to which it belongs; a careful inspection integrates all features in a manner difficult to analyze by measures. The observer himself is not always conscious of all the bases of his conclusion. The observer must use good judgment as to the definiteness with which the identification can be made from the features available; but good judgment is necessary in any case, whether the decision is made from the general appearance or from more objective measures. (Morgan, Keenan, and Kellman, 1943: 4).

This passage is vintage Morgan.[11] As a qualitative thinker in a field dominated by quantitative methods, Morgan could be savaged by insensitive colleagues who ridiculed his approach as 'celestial botany', but Morgan always considered himself as much an artist as a scientist. His method has proved the test of time. James Kaler (2002: 112), a leading expert on stellar classifications, has recently written:

> The standards become embedded in memory, and the typing of stars can proceed with impressive speed. There is a very important place for quantitative methods … Visual classification, however, is at present still useful in surveying in a reasonable amount of time the vast numbers of stars readily accessible to us.

One finds countless examples of Morgan's passion for visual pattern-recognition tasks in his publications, and especially in his personal notebooks. He had started to acquire art books in the 1930s and frequently commented on the works which captivated him. The following observations, written during the period when he was working on the Atlas, are typical:

> Sunset. May 19 [1942]. I want to be the man of the Rembrandt self portraits no. 40, 41, and 58. I want to look at women the way he looked at Hendrickje Stoffels and at the woman in no. 366. I want to look at the earth as Ma Yuan … and to feel like the sculptors of the Tang Bodhisattuas and to feel as does the head of Buddha. (Morgan, 1942; the numbers, above, refer to drawings in Dyke, 1927).

> Sunday afternoon. June 7 [1942]. Paraphrase of part of introduction of Cezanne book ... Elementary images can be created only by sacrificing the individual phenomena, the individual value of the human figure, the tree, the still-life subject. There is one characteristic of Cezanne's mode of representation which one may describe as aloofness from life, or better, as aloofness from mankind. In Cezanne's pictures the human figure often has an almost puppet-like rigidity, while the countenances show an emptiness of expression bordering almost on the mask. (Morgan, 1942).

Apart from the visual arts, Morgan's leisurely pursuits included putting puzzles together—he was famous for turning the colored sides of the pieces face down and assembling them from their shapes alone —and solving detective mysteries like G.K. Chesterton's Father Brown or Agatha Christie's Hercule Poriot stories. He was continually attentive to the patterns in the environment around Yerkes Observatory, which he documented in numerous photographs and paintings. The following passage, written at Walworth train station near Yerkes where he had gone to meet his daughter Emily (Tiki), is typical of countless passages. Awaiting her arrival he experienced—in a moment of Zen-like revelation—a world of profundities in the spare profiles of the telephone posts:

> Ah, another enchanted, cool-brilliant day; another communion; another sharpening of the senses, the vision, the physical response. How like delicate flower stems are these distant telephone posts. A progressive entrance into the world of reality – the World of the Self – during the past hour. Deeper and deeper, more and more removed from the ordinary. How far will it go? – how far can it go? There seems to be no limit in the Possibility – and no limit set by Time. (Morgan, 1963: 16 June).

## 7 ANOTHER CALIBRATION PROBLEM

In 1939, while using the 40-inch Yerkes refractor, Morgan realized that he could identify the different luminosity classes of B-type stars even from low-dispersion spectrograms. This was an important breakthrough since B-type supergiants, together with their brighter but rarer cousins, the O stars, are true stellar beacons, visible from relatively great distances. Morgan and others realized that these stars could, in principle, be used to map galactic structure provided one could calibrate the luminosities of the stars to their spectral types.

In principle, this is straightforward, but in practice difficult. The main problem is that, because there are so few of these stars—and none at all within a few hundred light-years of the Sun—they are all dimmed and reddened to some extent by interstellar dust, which is pervasive in the plane of the Galaxy. It exists as an omnipresent fog concentrated especially in the Galactic Plane in the dark clouds so well seen in Barnard's Milky Way photographs (as in Taurus, where the Pleiades illuminate some of the clouds at a distance of 400 light-years, and in Auriga and Perseus). Since the O and B stars, being young stars, are confined to the plane of the disk where the obscuration by dust is greatest, it turns out that even these luminous stars cannot be seen much beyond the nearest spiral features.[12]

It is possible to determine the amount of reddening of these stars even without knowing the detailed structure of the obscuring clouds. One must first calibrate the intrinsic color (the color of stars of a given spectral type independent of the effects of reddening). Then, since extinction of starlight by dust does not occur uniformly across the spectrum—it occurs about twice as efficiently at the blue end as at the red—by measuring the stellar magnitude in two different wavelength regions and taking the difference (i.e. the Color Index) one can, in principle, determine the degree of reddening of the star and so work out the effect of the dust.[13]

By correcting for the effects of extinction by interstellar dust and working out the distances of B stars in clusters, Morgan tried to calibrate the luminosity classes of his stars to their absolute magnitudes. His recognition, in 1939, that he could identify B stars even from low-dispersion spectra was important, because low-dispersion spectra could be obtained even for quite dim—thus far-away—stars. But he did not yet have any workable idea of how mapping these stars might lead to the discovery of the spiral arms. That recognition awaited developments from a totally unexpected line of research—Walter Baade's beautiful wartime work on the structure of the Andromeda Nebula.

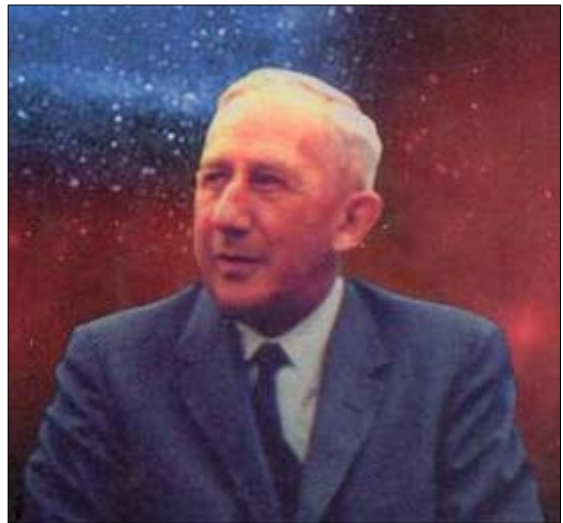

Figure 8: Walter Baade (after Osterbrock, 2001b).

## 8 THE TWO STELLAR POPULATIONS

In 1944, Walter Baade (Figure 8) published his seminal work on the two stellar populations, which turned out to be young and old stars (see Osterbrock, 2001b). Baade had come to the United States in 1931 with the intention of applying for citizenship, but he had lost his paperwork and never followed up on this. During World War II, he was classified as an enemy alien, which precluded him from taking part in war work. His unintended reward was to be given free rein with the 100-inch reflector when the lights of Los Angeles and Hollywood were blacked out, and by using remarkably fastidious observational techniques he was able to obtain deep plates which resolved the faint red stars in the nucleus of the Andromeda Nebula and its elliptical companions, M32 and NGC 205.

Baade submitted a paper describing this work to the *Astrophysical Journal* in 1944, and Morgan—who was assisting Struve in editing the journal at the time—immediately recognized its importance. He also saw that Baade's plates would not reproduce well and, ever the artist, succeeded in talking Struve into allowing him to make actual prints from Baade's negatives. In a labor of love reminiscent of E.E. Barnard's fussing over every photographic print in his *Atlas of Selected Regions of the Milky Way*, Morgan personally pro-duced and inspected prints of Baade's plates (Figure 9) and bound them into every individual copy of the *Astrophysical Journal* (which then enjoyed a circulation of between 600 and 800).

Baade's plates of the Andromeda Nebula showed clearly that in the spiral arms the hottest, most massive stars and open clusters were always associated with HII regions—diffuse nebulae of the Orion type, which had already, in 1939, been identified by Morgan's colleague, Bengt Strömgren, as regions of hot, ionized, interstellar hydrogen. The large complexes of nebulae and bright young O and B stars made up Baade's Population I. By contrast, the Galactic Nucleus and globular clusters were characterized by the fainter red giants of Population II. A crucial point, shown clearly in Baade's plates, was that concentrations of the O and B stars—the very same bright young hot stars that Morgan had been studying in our own Galaxy for several years—were the tell-tale markers that defined the spiral arms (e.g. see Figure 10). The reason they were concentrated in the arms was because they were young—necessarily so, since they were intrinsically bright, and would burn out before they had time to migrate very far from their place of formation.[14] This connection between stellar evolution and galactic structure was the essential clue that would ultimately produce the breakthrough leading to the recognition of our Galaxy's own spiral-arm structure.

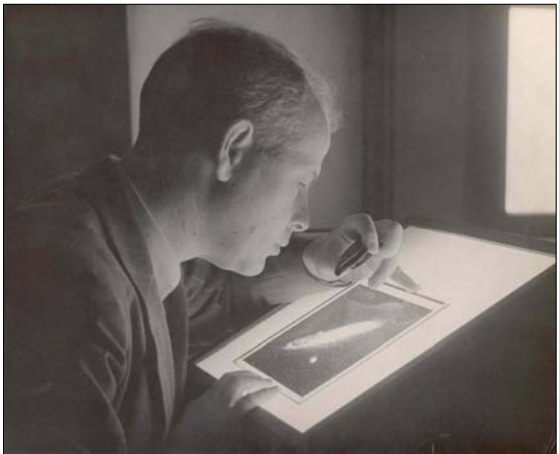

Figure 9: Morgan uses an ocular to inspect one of Baade's plates of the Andromeda Nebula (courtesy: Yerkes Observatory Archives).

As early as 1926, the two types of stars defined by Baade as Populations I and II had actually been recognized in our own Galaxy by Oort on the basis of their differing motions. Given their common interest in galactic structure, Baade and Oort began a vigorous correspondence shortly after the publication of Baade's 1944 paper. First Baade (1946) wrote to Oort on 23 September 1946:

> You mention in one of your remarks that the classical cepheids would be objects par excellence from which to determine the spiral structure. I think it is not certain yet that the longer period cepheids are especially concentrated in the spiral arms (they occur in the same regions in which the arms occur). But the B-stars of high luminosity are strongly concentrated in the spiral arms as my UV-exposures of the outer parts of M31 show most convincingly. I am therefore wondering, after reading Blaauw's fine paper about the Scorpius-Centaurus cluster, whether this extra-ordinary aggregation of B-stars is not in reality a short section of a spiral arm, the more so because in its orientation and motion it would fit perfectly into the expected picture (the arms trailing).

In the fall 1946 Oort gave a series of lectures at Yerkes Observatory which Morgan attended, and his pencil notes still exist, so that it is possible to follow Oort's reasoning on this important subject at just the time that Morgan was rapidly developing his own ideas on galactic structure.

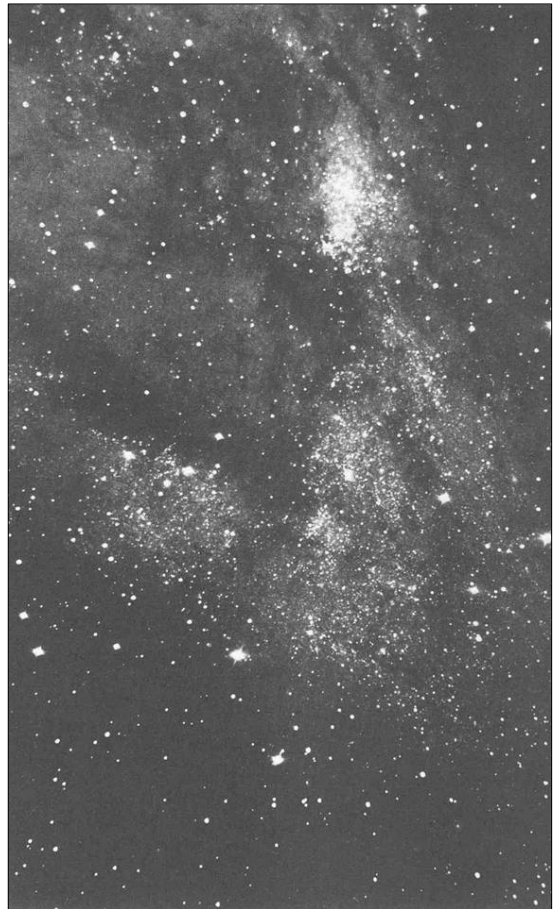

Figure 10: Baade's photograph of Population I objects in the Andromeda Nebula (after Baade, 1944).

Oort focused on the high-luminosity B stars, although in his Yerkes lectures he observed that "… one of the great difficulties … was that one did not know the point of the color excess. This was a consequence of the fact that … there are not many B-type stars nearer than 100 [parsecs], and even those are slightly colored." (Oort, n.d.). Morgan had, of course, been working on this very problem. In early 1947, Oort responded to Baade's earlier letter:

> I quite agree that a study of the early B-type stars would be one of the most important steps for finding the spiral structure of the Galactic System. I have been discussing this subject with Van Rhijn [Kapteyn's successor at the Kapteyn Astronomical Laboratory at Groningen] for some time, and when Van Albada left Holland in order to pass a year at Cleveland we suggested to him that he should try to start a program with the Schmidt camera for finding faint B-type stars in the Milky Way … This is a large programme, however, and I don't think the Warner and Swasey people are sufficiently interested yet to start it on a sufficiently big scale. How about future possibilities with the large Schmidt cameras on Mt. Palomar? (Oort, 1947).

Unbeknownst to Oort, Morgan had just teamed up with Jason Nassau from the Warner and Swasey Observatory on an ambitious survey to find B-type stars in our Galaxy. Morgan began to spend part of each year as a Visiting Professor of Astronomy in Cleveland, where he and Nassau identified the stars with the 24-inch Curtis-Schmidt camera at the Warner and Swasey Observatory. Later, Morgan used the 40-inch refractor at Yerkes Observatory to classify them rigorously by spectral type and luminosity. (This work would later be extended to more southerly regions by astronomers at Tonantzintla Observatory in Mexico.)

Morgan and Nassau's project was scarcely underway when, in December 1947, Baade spoke on the two stellar populations at an American Astronomical Society meeting at the Perkins Observatory in Ohio. By then it seemed increasingly likely that the spiral-arm structure of our Galaxy—if it existed—could best be mapped using the B stars, rather than by means of brute star-counts. Baade (1949a) later confided to Leo Goldberg that star-counts and statistical analysis had not led astronomers "… much beyond old William Herschel." Nassau and Morgan were of the same mind as everyone else, and fully expected that when they wound up their project of discovering the B stars they would have a good chance of working out the spiral-arm structure.

Nassau and Morgan were indeed finishing their project in the spring of 1949 when Morgan visited Pasadena and discussed with Baade the "… galactic survey for high-luminosity stars." (Morgan, 1949a). Shortly after their meeting, he wrote a long and important letter to Baade summarizing how far the project—and his thinking—had progressed by then:

When I came out, I had a fairly definite idea of what the galactic spiral structure within a radius of 3000 pc of the Sun is like; after the description of your own work I found that many of my ideas were wrong.

I regret very much that I did not take notes at the time of our discussion; as a consequence I do not remember the exact details of some important points …

(1) Were the large new emission nebulae which you have recently found in the region of the dark rift projected against the galactic center? ... I can't remember the approximate galactic longitudes.

(2) Have you also discovered similar nebulae [of] smaller dimensions in the anti-solar region? After thinking the matter over, it appears that the high luminosity stars which are observed within a fairly narrow range of true distance modulus[15] in the region of Cas[siopeia] and Per[seus] may well define a spiral arm located at a distance around 2-2.5 kpc. outside of the Sun. I have always been puzzled at the extent of the super giants surrounding the double cluster in Perseus; the concentration is probably explicable in terms of a spiral arm rather than as a physical cluster. In this respect, the region of Cepheus appears to be different in that high luminosity objects are observed over a greater range in the distance; this might be explained as a foreshortened effect for the outer spiral arm …

(3) Could the nearby extended dark nebulosity in Ophiuchus and diametrical[ly] opposite in Perseus and Taurus be considered the tattered outer remnants of the general extinction stratum of the spiral arm immediately within the position of the Sun?

It seems to me that within the next year it should be possible to reach a definite answer as to the location of the spiral condensations immediately within and without the position of the Sun. (Morgan, 1949b).

This letter has never been published, which is why I quote from it at length. It shows that Morgan now had the solution almost within his grasp. He had begun to pay attention to the distance moduli of the bright supergiant stars in Cassiopeia, Perseus, and Cepheus; moreover, he was already thinking of them as defining a spiral arm. He had also begun to pay attention to the distribution of the diffuse emission nebulae (HII regions). But though he was drawing close to the solution, it would be another two years before all those pieces, like the parts of a jigsaw puzzle or the characters and motives of a detective-novel, would finally and decisively fall into place.

Baade's response was delayed because he was then absorbed in trying to arrange the great sky survey using the 48-inch Schmidt telescope on Palomar Mountain (Edwin Hubble, who was originally charged with the program, had resigned owing to failing health). When Baade (1949b) did write, he agreed that Morgan was definitely on the right track:

Your interpretation of the large number of supergiants surrounding the double cluster in Perseus would be in line with the findings in the Andromeda nebula. There supergiants of very high luminosity are always bundled up in large groups which stand out as prominent condensations in the spiral arms.

The nearby extended dark nebulosities in Scorpius-Ophiuchus and Perseus-Taurus seem to be indeed manifestations of a single dark cloud ("streamer") which is tilted against the plane of the Milky Way and partly engulfs the solar neighborhood (both the Ophiuchus and the Taurus dark cloud are at a distance of only 100 parsecs).

The distribution of the B stars which you first pointed out to me is like this:

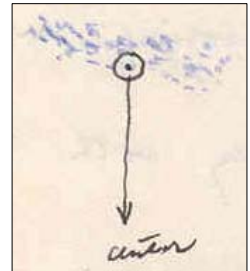

which leaves no doubt that the Sun is either *in* or close to the *inner* edge of the nearest spiral arm … I still think that the B star program will be the first to lead to definite information about spiral structure in our neighborhood and that you will push it as far as you can.

In July 1950, a symposium on galactic structure, led by Baade, was held at the University of Michigan Observatory. Morgan and Nassau were both there, and reported on the progress of their survey of the high-luminosity stars. Within a galactic belt 10° wide, they had identified 900 O and B stars. For most of these stars, the distances had not been determined, but for 49 OB stars and 3 OB groups Morgan and Nassau had been able to estimate distances (see Figure 11).

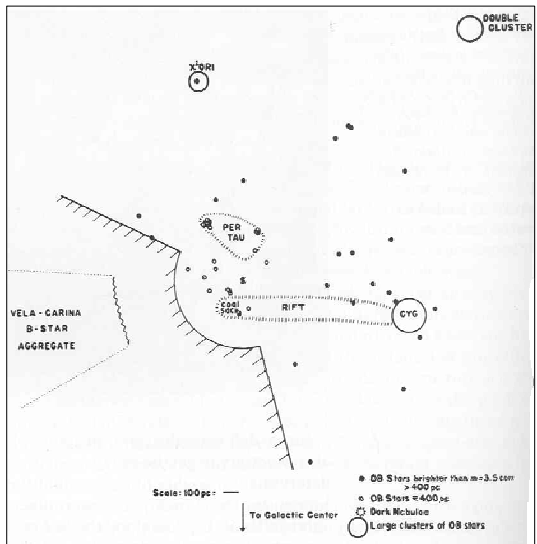

Figure 11: Plot showing the distribution of 49 OB stars and 3 OB groups with known distances (after Nassau and Morgan, 1951).

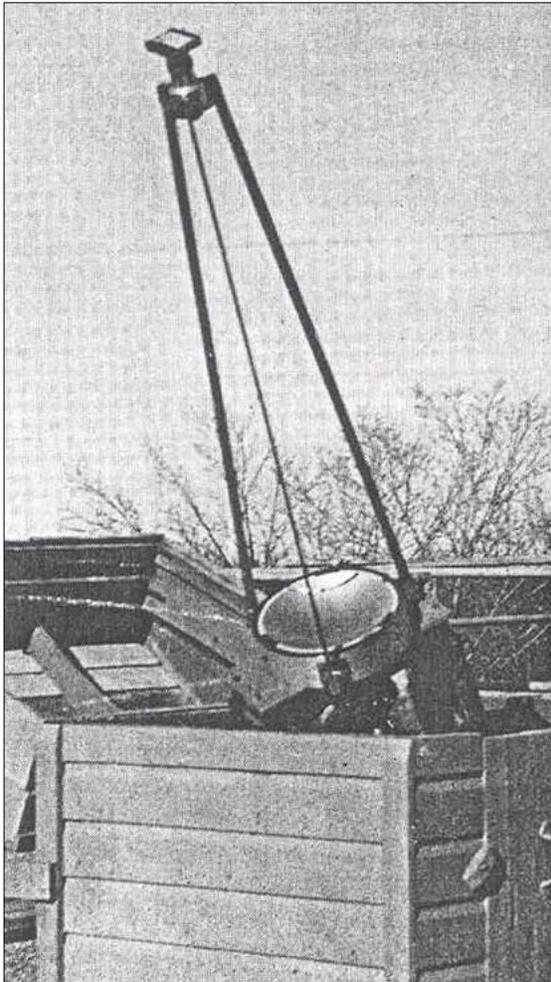

Figure 12: The Greenstein-Henyey Camera (courtesy Yerkes Observatory Archives).

According to a report on the symposium published in *Sky and Telescope* (1950), when Morgan and Nassau plotted these stars,

Combining the results with already existing knowledge of many facts about the galaxy and other galaxies, these astronomers suggested that the sun is located near the outer border of a spiral arm. The arm extends roughly from the constellation Carina to Cygnus. The fact that many faint and hence distant OB stars are found toward Cygnus indicates that we are observing the stars in the extension of this arm beyond the clustering in that constellation, that is, beyond 3,000 light years.

The part of the spiral arm near our sun contains a large cloud, or groups of small clouds, of interstellar dust and gas which obscures the distant stars and divides the Milky Way into two branches, easily visible to the naked eye. This obscuring cloud or rift is in the shape of a slightly bent cigar and is over 3,500 light years long. At one end of it is the southern Coalsack and at the other the brilliant group of OB stars of the Northern Cross … Dr. Nassau cautioned, however, that the evidence is insufficient to preclude the hypothesis that a great disorganization exists in the galaxy and that the star groupings do not trace definite spiral arms. (cf. Nassau and Morgan, 1951).

Nassau—and even Baade—had fully expected that a plot of B stars would furnish detailed information about our Galaxy's spiral structure, but they were deeply disappointed when nothing definite showed up from their plot, other than the well-known 'Gould belt', which represented the ring of bright hot stars close to the Sun that was originally mapped in the nineteenth century by the American astronomer, Benjamin Apthorp Gould.

## 9 EUREKA!

With the failure of this frontal assault on the spiral-arm structure, Morgan quickly regrouped. He now unfolded a grander strategy, which he hinted at in another paper presented at the same meeting. Innocently named "Application of the principle of natural groups to the classification of stellar spectra" (Morgan, 1951), its significance, indeed profundity, could hardly have been very apparent to anyone at the meeting. It was in fact as cryptic as the anagrams that earlier astronomers used to establish priority for their discoveries (yet at the same time effectively concealing them). In this paper, Morgan used the expression 'OB stars' to designate a category consisting of the O supergiant and early (young) B stars which formed what he called 'a natural group'. He noted that there was not much spread among the luminosity classes in type O, and even the early B stars showed only modest variations in luminosity. As he later explained later (Morgan, 1978), the significance was that it ought to be possible, "… by just a glance, [by looking] just a few seconds at each spectrum … to tell if a star was located …" in this rather narrowly-defined area of the Hertzsprung-Russell Diagram. Morgan (ibid.) felt that "… this was the crucial conceptual development." The stars in this region varied by only 1.5 or 2 magnitudes on either side of the means, which were around visual magnitudes −5 or −6.

Morgan was groping toward the concept of 'OB star associations' (although the term itself was not introduced until later by the Armenian astronomer, Victor Ambartsumian). The O and early B stars are found in loose aggregations typically of a few dozen stars (the majority of type B), which might be spread over a volume as small as an ordinary cluster or as much as a few hundred parsecs across. With a fair-sized group even of moderately discordant values of the luminosities, Morgan could pick the mean (around −5 or −6) and end up with a fairly reliable value for the group as a whole. Proceeding in this manner, he obtained good plots of their positions along the Galactic Plane, and this allowed him to reach out much further than Nassau had been able to do. Equally important, in Morgan's view, was the project he had intimated to Baade a year earlier: the identification of ionized HII regions, like the California Nebula close to Xi Persei, the so-called Barnard Loop in Orion, and the Rosette Nebula in Monoceros. They were, he recognized, completely analogous to the HII tracers of the spiral arms which Baade in 1944 had identified in the Andromeda Nebula. Inspired by Baade's photographs, Morgan combined his plots of OB associations and the HII regions of the Milky Way in a newly-energized and more focused attempt to trace the spiral arms.

At Yerkes there was at the time a wide-angle camera with a field of 140 degrees (see Figure 12). It had originally been developed during World War II by staff astronomers Jesse L. Greenstein and Louis G. Henyey for use as a projection system to train aerial gunners. However, it could equally well be used the other way around, as a camera, and under Morgan's direction two graduate students, Donald Osterbrock and Stewart Sharpless, began using it to photograph the Milky Way with narrow-band Hα filters (which had become available only after the War) in search of HII regions (Figure 13). Many of the HII regions were already well known; however, some were new, and because of the extraordinarily wide field of the photographs they were strikingly represented as the important extended objects they are (Morgan et al., 1952).

Sharpless had been one of Morgan's students, but Osterbrock's thesis adviser was Subrahmanyan Chandrasekhar, the master-theoretician whose approach was in many ways diametrically opposite to Morgan's. The theoretical astrophysicist Dimitri Mihalas, who later had an office across from Morgan and was befriended by him, has noted (pers. comm., November 2002) that Morgan and Chandra were like two mountain peaks —one was an observer and a pure morphologist, the other a mathematician and a master of theoretical deduction. Everyone else fell somewhere in the chasm between them. Young Don Osterbrock (Figure 14), through remarkable interpersonal tact and the astonishing versatility he later exhibited during a long and distinguished career as a research astronomer, administrator and historian of astronomy, was one of the few who managed to bridge that chasm.

The spiral structure of our Galaxy, if it existed, had proved to be much more difficult to recognize than anyone had ever imagined. It is in the nature of such things that it all seems perfectly clear in retrospect. In order to grasp just what was involved in making this discovery one must try to take oneself back in time. Morgan (1978) later recounted to David DeVorkin:

> One was looking at how these [OB stars] were and so on. Remember, there was nothing whatever known about the arms before. You have to remember this, because one goes back and thinks, well, you knew there was a tilt there … there were certain things at certain distances.

Although "Chance favors the prepared mind", as Louis Pasteur used to say, Morgan was hardly a 'sleepwalker', in Arthur Koestler's sense; he had been engaged in purposeful, goal-directed activity, following his hunch that plots of these highly luminous stars and the HII regions would finally led to the identification of the spiral arms. He had immersed himself in the problem for many years. But the discovery, when it came, came not as the result of the logico-deductive process; instead, he always insisted that it came in a flash—in a sudden dramatic moment of pattern-recognition.

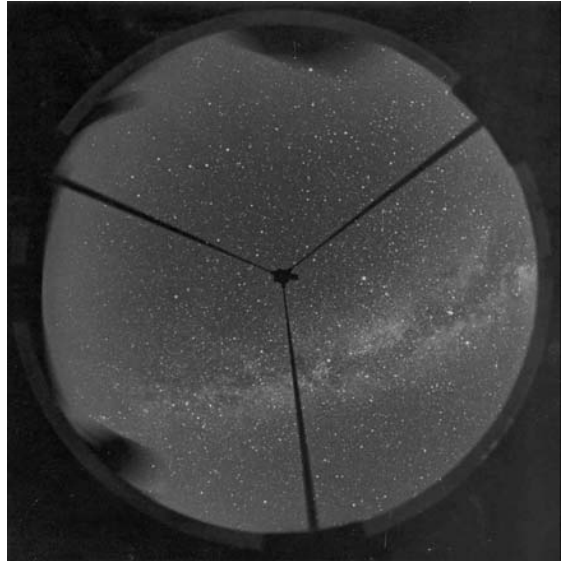

Figure 13: Representative image taken with the Greenstein-Henyey Camera (courtesy: Yerkes Observatory Archives).

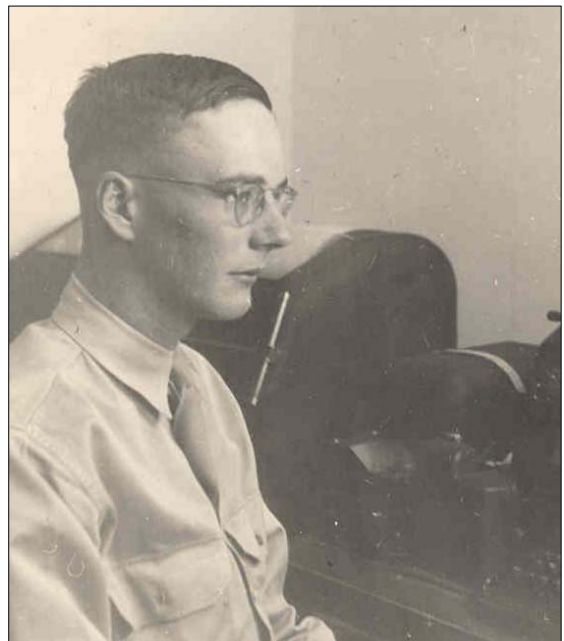

Figure 14: Don Osterbrock, during the War years, before entering graduate school (courtesy: Irene Osterbrock).

More than most astronomers, Morgan was receptive to the idea that the unconscious mind plays an important role in the discovery process. His personal notebooks are filled with reflections on psychoanalysis, and a number of passages allude specifically to the way he experienced the discovery of the spiral-arm structure in the fall of 1951 during what he described as his "… most creatively productive period … the two years centered on my 1952 breakdown." (Morgan, 1956: 9 December). Later, in this same notebook he wrote:

> December 29, 1956.
> Dear Book, what a strange thing the unbridled mind is. A sequence of thoughts can develop – move rapidly from stage to stage, and end in a conclusion (a definite, unique conclusion) in a few eye-closings. And what is the "unique conclusion" worth? Perhaps absolutely nothing. Conclusion may not result from premise; there may be spaces – infinities wide – between successive steps.

Morgan's most complete account of what happened that fall evening is given in an August 1978 American Institute for Physics interview with David DeVorkin, which is a singularly-valuable document (along with Morgan's various personal notebook entries on this subject) about the mysterious workings of a creative mind:

> This was in the fall of 1951 [he says elsewhere in the same interview that it was in October], and I was walking between the observatory and home, which is only 100 yards away [see Figure 15]. I was looking up in the sky … just looking up in the region of the Double Cluster [in Perseus], and I realized I had been getting distance moduli corrected the best way I could with the colors that were available, for numbers of stars in the general region … Anyway, I was walking. I was looking up at the sky, and it suddenly occurred to me that the double cluster in Perseus, and then a number of stars in Cassiopeia, these are not the bright stars but the distant stars, and even Cepheus, that along there I was getting distance moduli, of between 11 and 12, corrected distance moduli. Well, 11.5 is two kiloparsecs … and so, I couldn't wait to get over here and really plot them up. It looked like they were at the same distance … It looked like a concentration … And so, as soon as I began plotting this out, the first thing that showed up was that there was a concentration, a long narrow concentration of young stars … There are HII regions along there too … And that was the thing that broke [the problem] down. (Morgan, 1978).

This first spiral arm—the Perseus Arm—was traced between galactic longitudes 70 and 140 degrees (according to the system of galactic coordinates in use at the time).[16] As he plotted the OB stars, Morgan found out that in addition to this arm there was another, the Orion Arm, extending from Cygnus through Cepheus and Cassiopeia's chair past Perseus and Orion to Monoceros, i.e., between galactic longitudes 20° and 180 or 190°. The so-called Great Rift of the Milky Way marked a part of the inner dark lane of this arm; the Sun lay not quite at the inner edge but 100 or 200 light years inside it. It was the Sun's proximity to—indeed near-immersion in—this arm that had made its existence so difficult to identify.

There is no reason to doubt Morgan's account of that autumn night at Yerkes. As he walked from the Observatory to his home (and apparently right back again in order to do his plot), he experienced a 'revelation-flash', a moment of sudden pattern-recognition. As so often happens with those who have experienced a 'Eureka!' or 'aha!' experience (an insight-based solution to a seemingly unbeatable vexing problem), Morgan (1978) saw it as something ineffable, something that was impossible to define in words; it seemed to him to be an inspiration breaking through from the subconscious mind:

> The main thing that's of interest to me about this is that there was no syllogistic operation – given this, then this, and then this, and all that sort of thing. Nothing whatever. It was a flash. And this is the way things come, in flashes – everything I've ever been concerned with in discovery, has been a question of flashes. That doesn't mean one develops them. One had better get them down somewhere, if it's the middle of the night, or they're dead the next morning. You don't know that you have them.

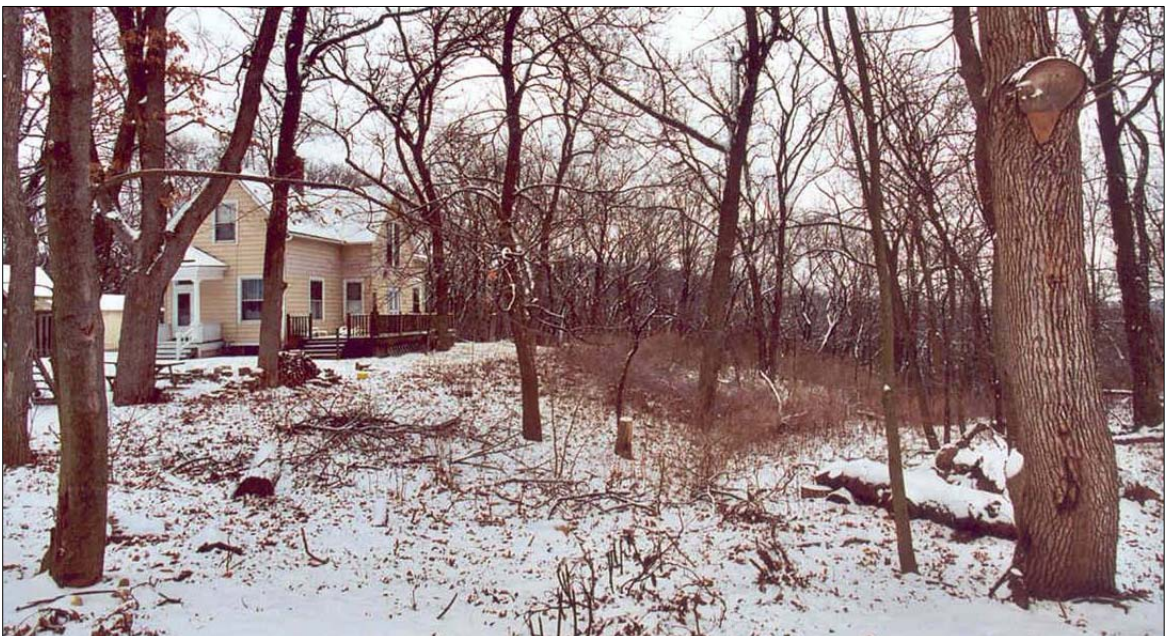
Figure 15: The Morgans' house on the grounds of the Yerkes Observatory (courtesy: Yerkes Observatory Archives).

Morgan, of course, had long demonstrated an unusual aptitude for pattern-recognition tasks, and had even based his MKK atlas of spectra on the human brain's marked ability to distinguish patterns. He himself had been born left-handed, but forced to learn to write with his right-hand (Mihalas, pers. comm., November 2002). He was, among astronomers, exceptionally artistic and highly creative. Neuropsychologically, he seems to have been either mixed-dominant or right-hemisphere dominant. Recent research on the psychology of insight by psychologists Edward Bowden of Northwestern University, Mark Jung-Beeman of Drexel University and their colleagues suggests that, although all thinking involves complementary right and left hemisphere processes, "… right hemisphere processing plays an important role in creative thinking generally and in insight specifically." (Bowden et al., 2005: 325).

It is certainly remarkable but entirely consistent with the literature on psychology of insight that even though Morgan had worked for years on the problem of the spiral-arm structure of our Galaxy (pursuing it through a systematic investigation involving a clear plan, implemented with meticulous attention to detail that began with identifying in low-dispersion spectra the distant O and B stars, continued through the correction of the effects of interstellar reddening, and culminated in his working out the luminosities, plotting the associations, and reinforcing the outlines of they defined with his map of the HII regions), that in the end the solution came to him in a flash, in a virtual eye-blink as the long-elusive embedded figure emerged, "… the flash inspiration of the spiral arms … a creative intuitional burst." (Morgan, 1956). As in the case of others who have experienced such insight-based solutions, Morgan "… experienced the solution as sudden and obviously correct (the Aha!) … [and] could not report the processing that had enabled him to reach the solution." (cited in Bowden et al., 2005: 323). As an artist, he was gratified that the resolution of his perplexity emerged from an inscrutable subconscious source.

Morgan's discovery was incarnated in a model in which old sponge rubber was used to depict the OB groups that he had identified (Figure 16). Later he added some concentrations of early B stars from the southern hemisphere (stars classified by Annie Jump Cannon as BO stars, those with hydrogen lines weak in the spectra that turned out to be a close approximation to Morgan's OB stars).

This more detailed scale model, constructed using balls of cotton, he presented in a slide at the American Astronomical Society in Cleveland, the day after Christmas 1951 (see Figures 17 and 18)—the meeting at which he received the ovation. The seats in the auditorium are located in banked rows that ascend from the stage, and the audience not only clapped their hands but they rose to their feet and started stomping on the wooden floor—in that acoustical space the effect was thunderous (David DeVorkin, pers. comm., July 2007). Since Oort, after introducing Morgan, had taken his seat, Morgan had nowhere to sit.

Morgan had finally received what he had always craved, the recognition of his peers. But within months he suffered a mental collapse; it was a "… complete personal crisis." (Morgan, 1978). During the spring after he discovered the spiral arms, he became depressed and unable to work, and his condition deteriorated to the point where he had to be hospitalized that summer. By the time he could return to work, the radio astronomers were rushing in and claiming enthusiastically that they had taken the mapping of the Galaxy so much further.

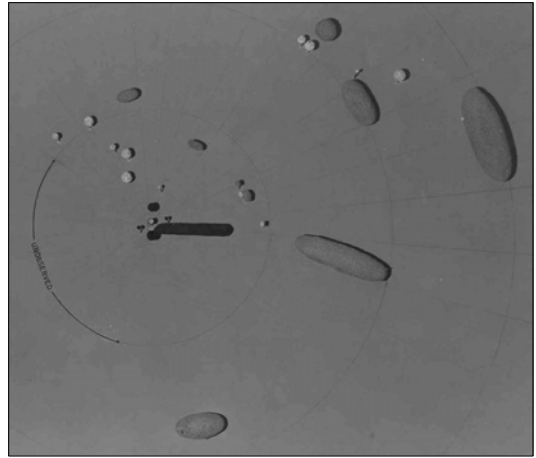
Figure 16: Morgan's sponge-rubber model (courtesy: Yerkes Observatory Archives).

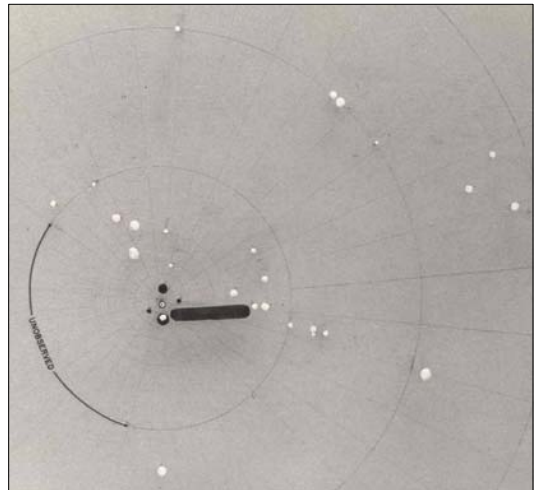
Figure 17: Morgan's later cotton-ball model and annotated diagram (courtesy: Yerkes Observatory Archives).

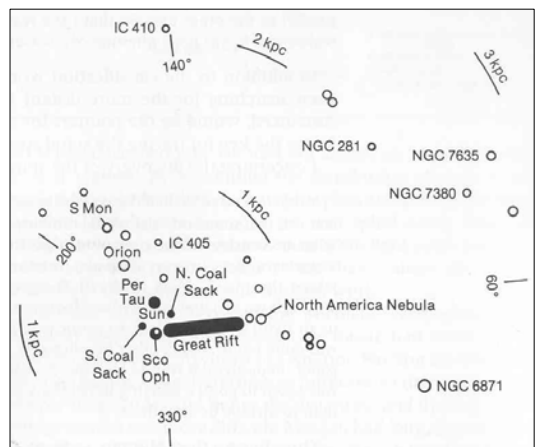
Figure 18: Legend to the model shown in Figure 17.

As he worked his way back to health—in part by means of those writing-exercises he committed to his notebooks—Morgan came to identify with Freud's self-analysis undertaken at the time Freud was making his most important discoveries in psychoanalysis. While reading the Freud-Fliess letters, he quoted Freud, who had said "… you can imagine the state of mind I am in—the increase of normal depression after the elation." To this, Morgan (1957b) added: "How true – how true! How often I have experienced this same phenomenon! Intrinsically, temperamentally, how similar I am to Freud!" In the same notebook, he later wrote: "Always there was melancholy in spring for me." (ibid.).

## 10 CONCLUDING REMARKS

One can speculate that if the radio astronomers had not 'stolen his thunder' then Morgan might even have won the Nobel Prize for his discovery. If he had, would it have resolved his struggle for self-esteem? Instead he continued, through a long and accomplished career, to grapple with the classification of galaxies and with the alternating creative phases of elation and let-down. His was the condition of many of those with creative temperaments, "… the greatness and misery of man …" as Pascal put it. Near the end of his life, Morgan (1983) wrote:

> A crucial conceptual breakthrough conversation with Osterbrock this morning. He would like to write my life … In the following conversation, I said I was not a genius; he said he was not sure I was right—that I had made "Conceptual Breakthroughs." The implication seemed to be that I might be … I told him that he had just given me the highest honor of my entire life.

In the discovery of the spiral-arm structure of our Galaxy, Morgan had achieved one of the most important scientific breakthroughs of twentieth-century astronomy and also caught an inspiration worthy of the great artists he so admired—those of the Trecento, the period from Cimabue to Giotto, who were the visual artists Morgan always thought had gone furthest in probing 'deepest reality'. Morgan (1956), like them—like Plato in the eternal realm of his universals—had secured his achievement in "… the hours of stillness – with supple brain – deep in the vistas of space, time, and form – that Heavenly World of Form."

## 11 NOTES

1. The only time this had happened previously was when V.M. Slipher announced the discovery of the large velocity-shifts of the nebulae at the A.A.S. meeting in 1913. Otto Struve (1953: 277) described the response to Morgan's paper as "… an ovation such as I have never before witnessed. Clearly, he had in the course of a 15-minute paper presented so convincing an array of arguments that the audience for once threw caution to the wind and gave Morgan the recognition which he so richly deserved."
2. When George Willis Ritchey finished working with the Mt. Wilson 60-inch Reflector in 1908, George Ellery Hale wrote in his *Annual Report of the Director* for 1909 that the observing program for the telescope was "… not yet definitely arranged." But Hale's plans would be decisively influenced by Kapteyn, who had sought the cooperation of major observatories in an observing program he called 'The Plan of Selected Areas', which aimed at nothing less than to determine the large-scale dynamics and structure of the Universe—a Universe which was then still thought by most astronomers to be a sidereal system bounded by the Milky Way. Kapteyn's plan called for the statistical analysis of results obtained from detailed surveys to be conducted in 206 'selected areas'—representative swatches evenly distributed around the sky. Hale was convinced, and argued that Kapteyn's work—especially his putative discovery that the stars moved in one of two opposing streams—bore directly on the problem which interested him most, that of stellar evolution. Although neither Kapteyn's 'star streams' nor his model of the Universe survived the test of time, the programs begun with the 60-inch telescope in the selected areas proved far-reaching, and much of the telescope's working life would be bound up in the great interwoven quests for the answers to stellar evolution and galactic structure. Kapteyn himself spent most summers as a Research Associate at Mt. Wilson from 1909 until 1914, advising Hale on the scientific course for the big telescope. Kapteyn was accompanied by his wife, and since women were not permitted to stay in the 'Monastery' (the main residence built for astronomers working on the mountain), Kapteyn lived in another cottage on Mt. Wilson; it is still known today as the 'Kapteyn Cottage'.
3. S Andromeda almost reached naked-eye visibility, and at the time it was assumed that it was like the ordinary galactic novae, which placed the Andromeda Nebula close by (see Jones, 1976). It was not until ordinary novae were finally identified in the Andromeda Nebula that astronomers realized the difference between novae and supernovae.
4. For biographical information on Morgan see Garrison (1995) and Osterbrock (1997).
5. In later years, Morgan sometimes tried to put a more positive 'spin' on his father's personality and approach to life. "My father, in a sense, was a very great man …" he told David DeVorkin (Morgan, 1987). "He told me once it took two generations to make a gentleman and he was the first. His father was the same kind of person he was. But it was very, very rough." (ibid.). He even dedicated his essay, "The MK System and the MK Process", in the proceedings of a workshop held in his honor at the University of Toronto in June 1983 "To my father, William Thomas Morgan (1877-??). You will never know what I owe you." (see Garrison, 1984: 18n). The use of ?? for the unknown date of his father's death strikes me as particularly poignant.
6. Morgan's first notebook was started in 1955, and the last (No. 247) was completed in 1990, by which time his thoughts were becoming scattered and somewhat random as he was suffering from Alzheimer's Disease. These notebooks are a unique resource, and document his mental life in almost Proustian detail. Of the 247 notebooks, 244 are at Yerkes Observatory. Jean Morgan, Morgan's second wife, after consulting with his closest and most trusted friends, Donald Osterbrock, Robert Garrison, and Dimitri Mihalas, wanted them to remain there and to be available to scholars. This was as she and they judged that Morgan himself would have wished. There are also a few earlier notebooks, which were not kept as part of this series, but which contain fascinating insights into his active interest in art as well as stellar classification in the early 1940s. Of those running continuously and in seriatim, the first begins on 20 April 1955, and the last, begun on 11 September 1990, peters out into rambling free-associations. The two earliest were removed; one was sent by Jean Morgan to a friend for consultation as to whether they might contain sensitive and highly personal materials. Another Morgan himself lost, and yet another one was given to extended family members and has not been available for study. The author has begun a close study of these notebooks as a step toward the goal of eventually producing a full-length biographical study of Morgan.

7. Frost's comment calls to mind Walter Baade's comments about Hubble's Ph.D. thesis, which, according to Osterbrock (pers. comm., March 2002), he called "… the most miserable thesis you ever saw."

8. The Balmer lines of hydrogen are, of course, very susceptible to Stark broadening, so the wings are a direct measure of the electron pressure in the stellar atmosphere. These lines are very strong in the B stars, and so the broadening can be detected at relatively low resolution. As the pressure is the weight of the overlying gas and the hydrogen is mostly ionized in the B stars, the stars with large radii (and hence luminosities) have the lowest pressures and the narrowest lines.

9. Note there is a difference between 'resolution' and 'dispersion'. The Mt. Wilson astronomers, with much larger telescopes, could afford to throw away most of the light on the spectrograph slit, thereby obtaining much better resolution at the same dispersion compared to those obtained with objective-prisms like the early Harvard spectra.

10. Although the reason that all the great spectral classifiers before Morgan were women was a result of Edward C. Pickering's scheme of having otherwise unemployed ladies do the routine work at the Harvard College Observatory—often for no pay at all—their aptitude may also have been, in part, a result of the general superiority of women over men for tasks such as facial-recognition (e.g. see Baron-Cohen, 2003). In this respect, the following notebook entry by Morgan (1957a), dated 1 January 1957, may be relevant: "My artistic sensitivity … may well have some personal feminine characteristics at its base."

11. But Andrew T. Young (pers. comm., 7 March 2007) glosses over it. He says: "I might add that I had a try at learning spectral classification myself. It is not at all an easy skill to pick up … So it's hardly true that 'anyone' can classify spectra—though the *Atlas* is certainly a big help … [and] the difficulty of spectral classification has turned out to be so severe that only a handful of people have become adept at it. For mass-produced luminosities, multicolor photometry has turned out to be the preferred way to go—along with the recognition that colors and MKK classifications don't match up, even for supposedly 'normal' stars."

12. It turns out that while the much fainter giants in globular clusters can be seen only a degree or so away from the Galactic Center in 'Baade's Window'—a real 'hole' in the dust, almost in the Herschelian sense—the O stars remain invisible at a fraction of that distance.

13. Many of the required photometric measurements had already been obtained by Joel Stebbins, C. Morse Huffer, and Albert Whitford at the University of Wisconsin. Stebbins and Whiford had devised a six-color spectrum, but Morgan, in collaboration with Harold L. Johnson, later invented the UBV system as a simpler version intended as an essential partner to the two-dimensional classification system. Their paper (Morgan and Johnson, 1953) is one of the most widely-cited papers in the general astronomical literature.

Though it had long been known that interstellar extinction, like atmospheric extinction, is 'selective'—that is, greater at shorter wavelengths—so that it produces reddening, there is not as strict a connection between spectral types and colors as everyone at first believed. There is a pretty tight correlation, but it is not perfect. The hope was that stars with identical spectral features would have identical colors, but this did not turn out to be true; the small but significant discrepancies are both puzzling and a considerable hindrance to getting accurate 'spectroscopic parallaxes' or photometric distances of individual stars. According to Andrew T. Young (pers. comm., 7 March 2007), the reason seems to be that "… colors depend mostly on sources of continuous opacity in the stellar atmospheres, but spectral types depend more on the line extinction coefficients, and these aren't as tightly coupled as one might hope, even for "normal" (Pop. I) stars; it's even worse for Pop. II, because the continuum opacity in cool stars is mainly the H-minus ion, where the electrons come from ionization of the metals; so less metals means less continuum opacity, so you see deeper into the star, and see more of the few metal atoms there. As a result, the spectra don't change nearly as drastically as the metal content does."

14. According to Andrew T. Young (pers. comm., 3 March 2007), "It is important to bear in mind that a star traveling at a speed of a kilometer per second will travel a parsec in a million years (very nearly). These young stars typically have speeds of only 10 or 20 km/sec. And, as their ages are generally on the order of 10 million years, they can travel only 100 or 200 pc before vanishing. Another essential fact is that even 20 million years is small compared to the timescale for the epicyclic motion about the local mean galactic rotation. The epicyclic period must be around 150–200 million years locally; divide by $2\pi$ to get the characteristic timescale, and you still have 30 million years or so—longer than all but the oldest B stars live."

15. The Distance Modulus is related to the distance of a star (d) in parsecs, and is given by the following formula, where $m$ is the apparent magnitude, and $M$ the absolute magnitude:
$$m - M = 5 \log d - 5$$

16. The old system, which measured the galactic longitude from one of the points where the Galactic Equator intersects the Celestial Equator—a choice that is purely arbitrary and without physical significance—has now been replaced by a new system using a slightly different Pole and measuring galactic longitude with respect to the Galactic Center (see Mihalas and Routly, 1968).

## 12 ACKNOWLEDGEMENTS

The author would like to thank Donald E. Osterbrock, Irene Osterbrock, Dimitri Mihalas, Robert F. Garrison, Andrew T. Young, C.R. O'Dell, Lew Hobbs, Ivan King, Owen Gingerich, David DeVorkin, Joseph E. Miller, Kyle Cudworth, Richard Kron, Judith Bausch, Richard Dreiser, Anthony Misch, Edward M. Bowden, and Mark Jung-Beeman for invaluable information and assistance. Institutional support was forthcoming from the Yerkes Observatory, the Mary Lea Shane Archives of the Lick Observatory at the University of California, Santa Cruz, the John Simon Guggenheim Memorial Foundation, and the Niels Bohr Library of the American Institute for Physics.

The late Professor Osterbrock—one of Morgan's collaborators in the discovery of the spiral arms—was particularly helpful, discussing Morgan's life and work with me on a number of occasions over many years. He also kindly made valuable unpublished material available to me. An early version of this paper was presented on the occasion of his Eightieth Birthday Celebration at the University of California, Santa Cruz. Unfortunately, he did not live to see the completion of this final version, or the full-length biography of Morgan which he hoped I would one day complete.

William Sheehan, M.D., a psychiatrist by profession specializing in functional brain imaging and the study of traumatic brain injury, is also an historian of astronomy who headed the organizing committee for astronomical history for the Donald E. Osterbrock Memorial Symposium at the University of California, Santa Cruz, in August 2007. He received a fellowship from the John Simon Guggenheim Memorial Foundation in 2001 for his work on the structure and evolution of the Milky Way, and won the gold medal of the Oriental Astronomical Association in 2004 for his prediction and confirmation of flare events on Mars. He is a member of IAU Commission 41 (History of Astronomy). Among his many books are: *The Immortal Fire Within: the life and times of Edward Emerson Barnard* (1995), *The Planet Mars* (1996), *In Search of Planet Vulcan* (with Richard Baum, 1997), *Epic Moon* (with Thomas A. Dobbins, 2001), and *Transits of Venus* (with John Westfall, 2004). He has published over 150 papers and articles on the history of astronomy, many of them in *Sky & Telescope*, where he served for many years as a Contributing Editor. The IAU has named asteroid 16037 for him.